\documentclass[a4paper]{article}
\usepackage[T1]{fontenc}
\usepackage[utf8]{inputenc}
\usepackage[italian,english]{babel}
\usepackage[autostyle,italian=guillemets]{csquotes}
\usepackage{multicol}
\usepackage{microtype}
\usepackage{guit}
\usepackage{booktabs}
\usepackage{graphicx}
\usepackage{mathptmx}
\usepackage{amsfonts}
\usepackage{amssymb}
\usepackage{amsmath}
\usepackage{ifpdf}
\usepackage{dcolumn}
\usepackage{booktabs, longtable}
\usepackage{hyperref}
\usepackage{textcomp}
\usepackage{tabularx,array}
\usepackage{ulem}
\usepackage{placeins}
\usepackage{epstopdf}
\usepackage{etex}
\title{New Meteorological and Geological Study \\ of Taviano (LE)} 
\author{Tasselli, D. \\ TS Corporation Srl - Astronomy and Astrophysics Department\\ Regione Salamia, 10010 Andrate TO - Italy \\ E-mail:diego.tasselli@tscorporation.org \\ \\ Ricci, S. \\ TS Corporation Srl - Meteorogical and Climatic Change Department\\ Regione Salamia, 10010 Andrate TO - Italy \\ E-mail:stefano.ricci@tscorporation.org \\ \\ Bianchi, P. \\ TS Corporation Srl - Geology and Geophysics Department\\ Regione Salamia, 10010 Andrate TO - Italy \\ E-mail:pamela.bianchi@tscorporation.org}
\date{}
\begin{document}
 
 \maketitle
 
 \begin{abstract} 
This paper contains the result of the elaboration of informations about Solar Irradiation, Geological, Meteorological and Climatic from the point of view of the quantitative data and social interaction recorded in Taviano (LE) over 24 years. These data are compared to check local variations, long term trends, and correlation with mean annual temperature. The ultimate goal of this work is to understand long term climatic changes in this geographic area. The classes of event considerated are hydro-geological phenomena, sun irradiation, seismic, volcanic, meteorological and climatological event. Only event occurred between 1990 and 2014 are considerated. The analysis is performed using a statistical approach. A particular care is used to minimize any effect due to prejudices in case of lack of data. Finally, we calculate the annual average from the monthly ones. Data on this paper don't come from a complete census of phenomena; they are considered enough representative of the accepted vulnerability level at the beginning of this study.
 \end{abstract}

{\footnotesize This paper was prepared with the \LaTeX \\}
\begin{multicols}%
{2}
\section{\normalsize Introduction} In this paper we present for the first time an analysis of geological, meteorological and climatic data revealed in Taviano (LE) by ``SMCS - Stazione Meteo-Climatica e Sismologica'' a project by Meteorological and Climatic Change and Geological Department of TS Corporation Srl.\cite{stazionemeteo:2011ug} compared in order to check local variations or meteorological conditions. We discuss the geological data, annual temperature means and differences between day time and night time mean values and their comparison with the down time. \subsection{\normalsize Location} Taviano is a little town in the south of Italy. Here are identifiable altimeter data, the geographic coordinates and seismic data.\\ \\
\begin{tabular}{c|c|c}
\hline
\scriptsize \textbf{Latitudine} & \scriptsize \textbf{Longitude} & \scriptsize \textbf{Share} \cr
\hline
\scriptsize $39^\circ 59' 4,20$'' N & \scriptsize $18^\circ 5' 15,36$'' E & \scriptsize 57 mt \cr
\hline 
\end{tabular}  \\  \\ \scriptsize {\bf {Geographic Data of Taviano}} \\ \\
\normalsize  The geological map of the town is shown in Geological and Seismic data section. \cite{ISPRA:2015} \\

\begin{tabular}{lp{0.2\textwidth}}
\hline
\scriptsize \textbf{Seismic Zone} & \scriptsize \textbf{Description}  \cr
\hline 
\scriptsize 4 & \scriptsize Area with very low seismic danger. is the least dangerous area, where the possibility of seismic damage is low. The area has a value of ag < 0,05g. \cr
\hline 
\end{tabular} \\ \\ \scriptsize {\bf {Seismic Data of Taviano}}
\subsection{\normalsize Mancaversa beach} \normalsize Taviano has a portion of its territory that borders the Ionian Sea. The place called "Marina di Mancaversa" is located on the Ionian coast about 5 km from the center of Taviano. The coast is mostly rocky and 900 meters long.
\section{\normalsize Annual data analysis} \normalsize The following table identifies climate data assigned by Decree of the President of the Republic n. 412 of 26 August 1993.\cite{Tasselli:2011ug}. \\ \\
\begin{tabular}{c|c}
\hline
\scriptsize \textbf{Climatic Zone} & \scriptsize \textbf{Day Degrees} \cr
\hline 
\scriptsize C & \scriptsize 1.099 \cr
\hline 
\end{tabular} \\ \\ \scriptsize {\bf {Climatic Parameter of Taviano}}
\section{\normalsize Meteo-Climatic Parameter} 
\normalsize In this section we describe air temperatures (T), Dew Point, Humidity, Pressure, Day Time and Night Time Variation, Rain's Days and Fog's Day, obtained by an accurate analysis of the meteorological data from local data by archive \cite{datimeteo:2013} \cite{eumetsat:2013}. \\ Should be noted that the values considered are related to the last twenty-four-year average and made available for the period 1990-2014. \\ \\
\begin{tabular}{lp{0.10\textwidth}} 
\hline
{\footnotesize Average Annual Temperature}&  {\footnotesize $16,76 ^\circ C$}  \cr
{\footnotesize T average warmest (Jul-12)}& {\footnotesize $28,16 ^\circ C$} \cr
{\footnotesize T average coldest (Feb-12)}&  {\footnotesize $4,43 ^\circ C$} \cr 
{\footnotesize Annual temperature range}&  {\footnotesize $9,90 ^\circ C$}  \cr
{\footnotesize Months with average T > $20 ^\circ C$} & {\footnotesize  104} \cr
{\footnotesize Total rainfall 1990-2014 [mm]}& {\footnotesize 16003,70} \cr
{\footnotesize Rain Days }&  {\footnotesize 1092}  \cr
{\footnotesize Fog Days }&  {\footnotesize 770}  \cr
{\footnotesize Storm Days} &  {\footnotesize 188}  \cr
{\footnotesize Rain/Storm Days }& {\footnotesize 595}  \cr
{\footnotesize Rain/Snow Days}&  {\footnotesize 11}  \cr
{\footnotesize Rain/Fog Days} & {\footnotesize 45}  \cr
{\footnotesize Rain/Thunder/Fog Days} & {\footnotesize 31}  \cr
{\footnotesize Snow Days} & {\footnotesize 11}  \cr
{\footnotesize Wind Speed max Km/h (Mar-2002)} & {\footnotesize 43,45}  \cr
{\footnotesize Wind Speed min Km/h (Nov-2010)} & {\footnotesize 6,00}  \cr
{\footnotesize Rain max mm (Oct-2004)} & {\footnotesize 343,00}  \cr
{\footnotesize Rain min mm (Aug-2000)} & {\footnotesize 0,00}  \cr
{\footnotesize Earthquake Min (2011/12/20)} & {\footnotesize 0,5 Mw}  \cr
{\footnotesize Earthquake Max (2006/07/06)} & {\footnotesize 2,4 Mw}  \cr
{\footnotesize Earthquake Deep Min (2012/08/21)} & {\footnotesize 5 Km}  \cr
{\footnotesize Earthquake Deep Max (2010/04/28)} & {\footnotesize 10 Km}  \cr
\hline
\end{tabular} 
\\ \\ \mbox{\bf{\footnotesize  Parameter of this Study}}
\subsection{\normalsize Solar Radiation Territory}  \normalsize The data irradiation of territory taken from the parameters and the data prepared by the European Union, demonstrate the trend of irradiation for Taviano, visible in next table: \\ \\
\begin{tabular}{|c|c|c|r|}
\hline
     \bf {\footnotesize Month} & \bf {\footnotesize DNI}& \bf {\footnotesize Month} & \bf {\footnotesize DNI}\cr \hline
       {\footnotesize Jan} &  {\footnotesize 3310} & {\footnotesize Feb} &  {\footnotesize 3980} \cr
       {\footnotesize Mar} &   {\footnotesize 5350} & {\footnotesize Apr} &  {\footnotesize 6390} \cr
       {\footnotesize May} & {\footnotesize 7570} & {\footnotesize Jun} &  {\footnotesize 9040} \cr
       {\footnotesize Jul} &   {\footnotesize 9410} &  {\footnotesize Aug} & {\footnotesize 8680} \cr
       {\footnotesize Sep} & {\footnotesize 6470} &  {\footnotesize Oct} & {\footnotesize 5100} \cr
      {\footnotesize Nov} & {\footnotesize 3820} & {\footnotesize Dec} &  {\footnotesize 3220} \cr \hline
     \bf {\footnotesize Year} & \bf {\footnotesize 6040} & & \cr
\hline
\end{tabular}
\\ \\ \mbox{\bf{\footnotesize Direct Normal Irradiance (Wh/m$^2$/day) }} \\
\normalsize The weather data and the graphs show extrapolated for the territory covered by the study, including a radiation in the range between 1350 and 1400 kWh /1kWp as map prepared by the European Union \cite{UE:2013} and visible in figure 19, characterized in over the months, irradiation presented in the graph in Figure 19, which shows the data of the table above, which shows the territory of Taviano, a total irradiance Annual of 6040 Wh/m$^2$/day.
\subsection{\normalsize Temperature} \normalsize In this section we describe air temperatures (T) obtained by an accurate analysis of the meteorological data. \\ The average temperature shows a tendency to intersperse over 4 years, an increase and a decrease in the average
the study period 1990-2014. \\ The period of greatest temperature increase has occurred since 2010, with an increasing trend of about $1^\circ$, while the greater decrease for the period of study has taken place in 2002. \\ At the turn of these periods, the study highlights a trend in rising and then held constant for the period between 1996 and 2001, with a trend down slightly in 2004-2006. Next graphics show this evidence: \\ \\
\includegraphics[width=0.49\textwidth{}]{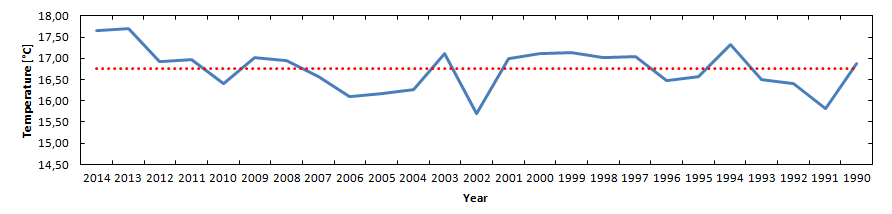} 
{\scriptsize \bf \textit{\textit{{\footnotesize Relationschip of Year Temperature and Average 1990-2014}}}}. \\
 \\The values are calculated considering the entire measurement period (1990-2014), drawing on data from the Annuals published by "Il Meteo.it" \cite{datimeteo:2013} and Eumetsat  \cite{eumetsat:2013} for the period 1990-2014. \\ \\ \includegraphics[width=0.49\textwidth{}]{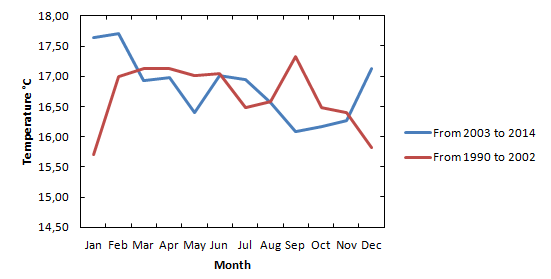} 
 {\scriptsize \bf \textit{\textit{{\footnotesize Variation of Month Temperature about year 1990-2002 and 2003-2014}}}} \\
 \\ The graphs shown in Figures 4 and 5 show the performance of the maximum and minimum temperatures proportional to the average of the period of study (1990-2014). The study shows the following trends:
\begin{itemize}
\item Minimum temperatures: there was an increase of the values recorded with a higher peak in February 2012, with $4,43^\circ C$;
\item Maximum temperatures: evidence for the period 2007-2009 is an increasing trend compared to the 1990-2014 average, while evidence a decrease in the period 1990-1997. The maximum temperature was $28.16^\circ C$ in July 2012.
\end{itemize} 
Table 5 shows the trend of the points of maximum temperature expected in the summer quarter (June-July-August) and are shown in Figure 15. Study shows that the month of July 2012 recorded the highest average maximum value in a context in which the entire month recorded in the period 1990-2014 values always above average. \\ Alway Table 5 shows the trend of the points of minimum temperature expected in the winter quarter (January-February-March) and are shown in Figure 16. \\ Study shows that the month of February recorded the highest average minimum value, in a context in which the entire month recorded in the period 1990-2014 values always above average. \\ Studying in particular the months with the values of minimum temperature and maximum minors, the following is noted:
\begin{itemize}
\item the month of July (characterized by the presence of the highest value observed in maximum temperatures), shows that trends in temperature has been getting consistently below average for the period 1990 to 2014, while there were two exceedances of this value over the years 2012 and 1998, in agreement with what evident from the graph in Figure 13;
\item the month of February (characterization from the value of the lowest minimum temperature for the period of study), shows that the trend of temperatures has always been, in agreement with what reported in the graph in Figure 14.
\end{itemize}
\subsection{\normalsize Dew point} In this section we describe Dew point obtained by an accurate analysis of the meteorological data. This study showed: 
\begin{itemize}
\item an increase in the year: 1992, 1993, 1994 (the hightest Dew Point period for this study), 1995, 1998, 1999, 2013 and 2014;
\item a decrease in rest of study period, with the lowest Dew Point in 2008 year. 
\end{itemize}
Graphics in Figure 11 show this trend. 
 \subsection{\normalsize Humidity} In this section we describe humidity obtained by an accurate analysis of the meteorological data. 
 The study highlights a gap in the annual humidity values equal to 69,38\%, calculated according to this formula:
\begin{equation}\frac{av}{am}\end{equation} 
{\footnotesize \textit{ {\bf av} = annual value, {\bf am} = average moisture 1990-2014}} \\ \\ The graph in Figure 9 shows a trend tends to be stable in the values obtained with threshold in growth over the period 1990,1991, 1992,1993,1994 (the hightest Humidity point for this study),1995,1996 and 2014, in rest of study period, with the lowest in 2008 year.
\subsection{\normalsize Pressure} \normalsize In this section we describe pressure obtained by an accurate analysis of the Pressure data. \\ The analysis of the data showed:
\begin{itemize}
\item an increase in atmospheric pressure over the years: 2000, 2004 (year with the highest level of pressure throughout the study period), 2007, 2008, and 2011;
\item a decrease in atmospheric pressure in rest of study period, with the lowest data in 2010 (the year with the last level of pressure throughout the study period).
\end{itemize}
In Figure 10 we can see the Pressure Graphics for this study. 
\subsection{\normalsize Day time and night time variation} In this section we describe number of Day time and night time variations obtained by an accurate analysis of the meteorological data. 
The annual averages of the differences between day time and night time temperatures $\Delta T$ have been computed and the results are reported in Table 2. Also in Figure 3, we can see the plot of oscillations of the $\Delta T$ seem to reduce the amplitude during the years. \\ The difference of temperatures, with an average difference of $9,90^\circ C$ (see table 2). Table 2 and Figure 3 show these effects.
\subsection{\normalsize Rain's Days} In this section we describe number of Rain's days obtained by an accurate analysis of the meteorological data. \\  An increase in extreme weather events and abnormal, leading to a potential increase in precipitation intensity for each event, especially in areas where there is an increase of average precipitation. \\ \includegraphics[width=0.49\textwidth{}]{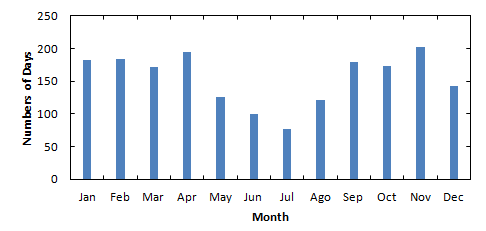} 
{\scriptsize \bf \textit{\textit{{\footnotesize Days Rain for Month from 1990 to 2014}}}}. \\ \\
 The study period showed an abnormal increase in rainfall. The pie chart shows the ratio between the total mm of rain measured in a year and the total rainfall measured in the period 1990-2014, according to this formula: \begin{equation}Total Rain=\frac{a}{b}x100\end{equation} 
{\footnotesize \textit{{\bf a} = annual rainfall value, {\bf b} = total value of the rain period}} \\ \includegraphics[width=0.49\textwidth{}]{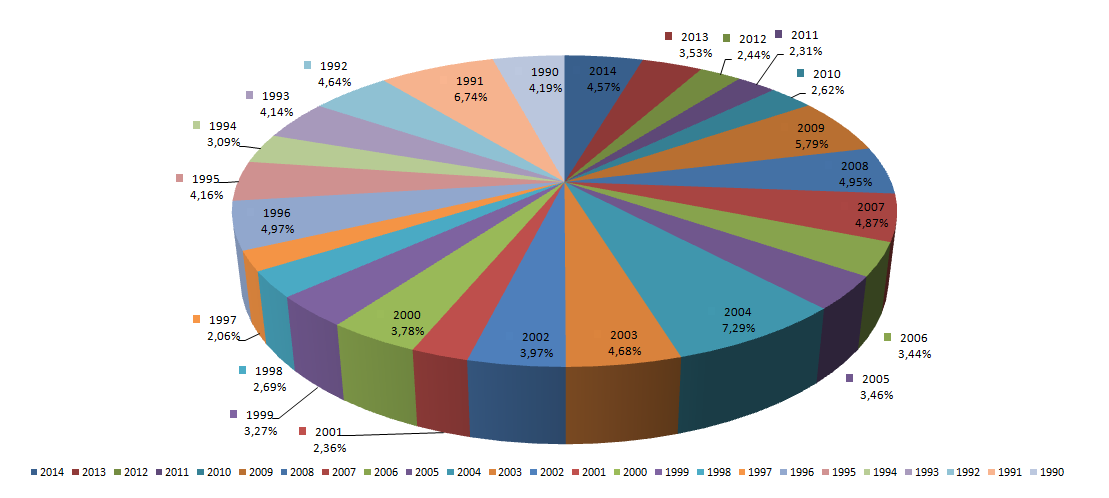}
{\scriptsize \bf \textit{ \% of Rain for this study}}.
\\ \\ In Figure 6 and 7 we can see the total of rain/year in this study. The values are visible in the table 3. Next pie chart evidence type of rain for this study. \\ \\
\includegraphics[width=0.5\textwidth{}]{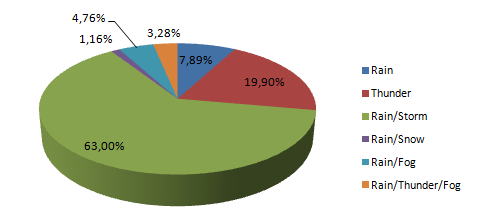}
{\scriptsize \bf \textit{Type of Rain for this study}}. \\ \\ Figure 17 evidence the total rain in mm, for only month. \\ The study shows an abnormal increase in rains, shown in Figure 7, in the years: 2014, 2009, 2008, 2007, 2004 (the highest year period mm of rain fell), 2003, 1996, 1992, 1991 , with an average increase of 134.71\% of the total annual rainfall mm which is equal to 640.15 mm on average for year.
\subsection{\normalsize Fog's days} In this section we describe number of Fog's days obtained by an accurate analysis of the meteorological data. \\ The study showed a trend increase in the presence of days with fog and evidence that January is the first month for number of Fog's day and June and July are the laster. Figure 8 we can see the number of Fog days by this study. \\ \\
\includegraphics[width=0.5\textwidth{}]{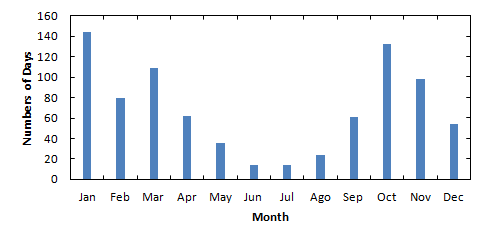}
{\scriptsize \bf \textit{Days Fog for Month from 1990 to 2014}} 
\subsection{\normalsize Wind speed} 
\normalsize In this section we describe the wind speed. \\ In the above image you can see the map of wind speed insistent on the territory of Taviano, as seen from the map generated from Atlas wind \cite{atlantevento:2013}. \\ The study of daily wind speed has allowed to estimate on a monthly basis throughout the period included in this study:
\begin{itemize}
\item a decrease in the average for the period 1990-2014 in the speed of the winds, with values of 6 Km/h in November 2010;
\item an increase with higher gusts, in high winds, especially in the period between the autumn and winter, with values of 43,45 Km/h in March 2002.
\end{itemize} 
\normalsize The study also shows that the month of March is the one with the trend towards greater variation in the wind speed, as well it appears from the comparison chart, maximum and minimum wind speed for this study shown in Figure 12. \\ \\
\includegraphics[width=0.25\textwidth{}]{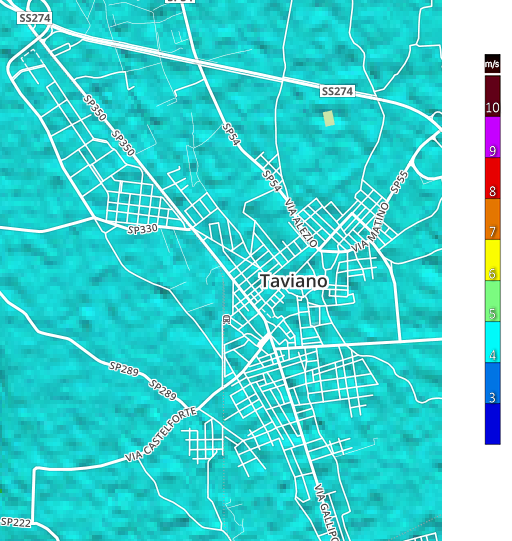} \\ 
{\scriptsize \bf \textit{Wind Speed Atlant of Taviano \cite{atlantevento:2013}}}  
\section{\normalsize Geological and Earthquake data} In this section we describe the Geological and Earthquake data. \\ The reconstruction of the geologic and lithological was made according to the study of aerial photographs, the interpretation of the stratigraphic wells, and finally on the basis of a detailed geological survey.  The town of Taviano is placed at a modest slope topographic oscillating for $0 ^\circ$ of the seaside town of Mancaversa, to $58 ^\circ$ of the municipality itself. \\ The current geological configuration was formed by tectonic distension that affected the basement carbonate during the service sector and that has created a series of depressions, where Pleistocene sedimentary sequences currently present, have been placed in succession. The formations are present consist mainly of Plio-Pleistocene sandy-arenaric and/or calcarenitic, resting on clay  deposits.
\subsection{\normalsize Description of the main outcrop formations}
The rocks outcropping in the territory of Taviano are:
\begin{itemize}
\item Limestones Melissano: this training which is the base on which rest the next, shows stratification variables to undulating, with subvertical fractures, with diaclase and leptoclasi structures with physical and mechanical secondary due to the action of karst. The lithology shows a brown or hazel color, compact, in layers and benches, alternating levels of gray or hazel;
\item Calcareniti of Salento: this training is based on the previous, derives from the accumulation of materials derived from erosion of the greenhouses of which was made the previous base. The lithology shows a gray yellow light and compact, coarse limestone and calcareous sabbioni, more properly defined in the terminology "Tufi".
\item Formation of Gallipoli: the formation of Gallipoli is constituted by two fundamental rock types which are:
\begin{itemize}
\item The marl clay at the base: they have a bluish-gray tint, are less stratified and contain varying percentages of fragments quazo sharp edges;
\item The sandy marl at the top of the above: constituent layers well defined, have a yellowish tint or gray-yellowish, have a clay content, and mainly consist of fragments of quartz grain medium / fine.
\end{itemize}
This formation is constituted by two lithological units that are sands and clays outcropping gray-blue, present in depth. The clays in question, correlate well, both from the point of view lithological that stratigraphic, subappenine Plio-Pleistocene clays or clay gray-blue Calabriane, found in different areas of Puglia, from The board at the end murgiana Fossa, the Murge and the Salento. Basically the characters of these clays are largely comparable along all areas of outcrop. \\ Stratigraphically, they are placed in the middle part of the sedimentary cycle Pliocene-Pleistocene. They are based, in continuity of sedimentation on calcareniti pliopleistoceniche (Calcareniti Gravina). Pass upwardly gradually, in general, to deposits sandy or calcarenitic calabriani, constituting the terms of closure of said Cycle.
\end{itemize}
\subsection{\normalsize Earthquake}
\normalsize This study evidence an stability of Earthquake activity for Taviano. \\ In next figure we can see the number of Earthquake in Taviano, and table evidence the number of Earthquake event by year. \\ \\
\includegraphics[width=0.45\textwidth{}]{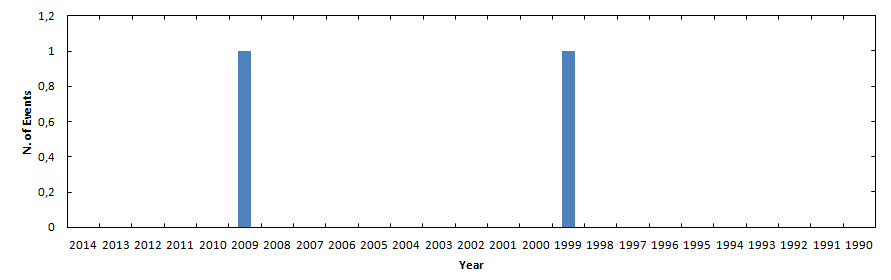} \\
{\scriptsize \bf \textit{Number of Earthquake for 1990 to 2014}} \\ \\
\begin{tabular}{|r|c|c|c|}
\hline
\scriptsize \textbf{Year} &\scriptsize \textbf{N. of Events} &\scriptsize \textbf{ Year} & \scriptsize \textbf{N. of Events}\\
\hline
\scriptsize 2009 & \scriptsize 1 & \scriptsize 1999 & \scriptsize1 \\
\hline
\end{tabular} \\ \\
\scriptsize {\bf {Number of Earthquake in Taviano by Year}} \\ \\
\normalsize The study evidence that the 2 Earthquake are production on deep from 5 to 10 Km. Next graphics evidence this. \\ \\
\includegraphics[width=0.5\textwidth{}]{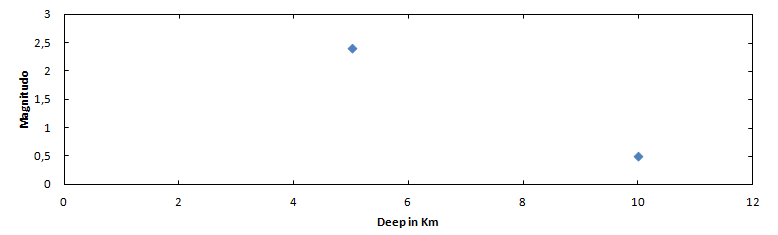} \\
{\scriptsize \bf \textit{Magnitudo/Deep for 1990 to 2014}}
\section{\normalsize Conclusion} 
We presented for the first time an analysis of geological and longterm temperature data directly obtained from Taviano local meteorological site, inside urban concentration and well above the inversion layer. \\
From a meteorological perspective Taviano falling within the territory of the southern Salento which it has a Mediterranean climate, with mild winters and summers warm moist. \\ According to the reference averages, the average temperature of the month coldest, February, amounts to about $+4,43 ^\circ C$, while that of the warmest month, July, is about $+28,16 ^\circ C$. \\ Average annual rainfall, who prowl around 640,15 mm, have a minimum in the spring-summer and a peak in autumn winter. \\ Based on data on wind, weakly affected western currents thanks to the protection given by the greenhouses salting that create a system to shield. Instead the current autumn and winter by South-East, aimed in part the increase in precipitation as well highlighted in the table of rain visible in section 3.7.
\section{\normalsize Acknowledgments}
We would like to thank Silvia Gargano for his helpful suggestions, for his professionalism and for the support to make the paper more complete.
\end{multicols}
\newpage
\begin{figure}
\begin{center}
\includegraphics[width=0.8\textwidth]{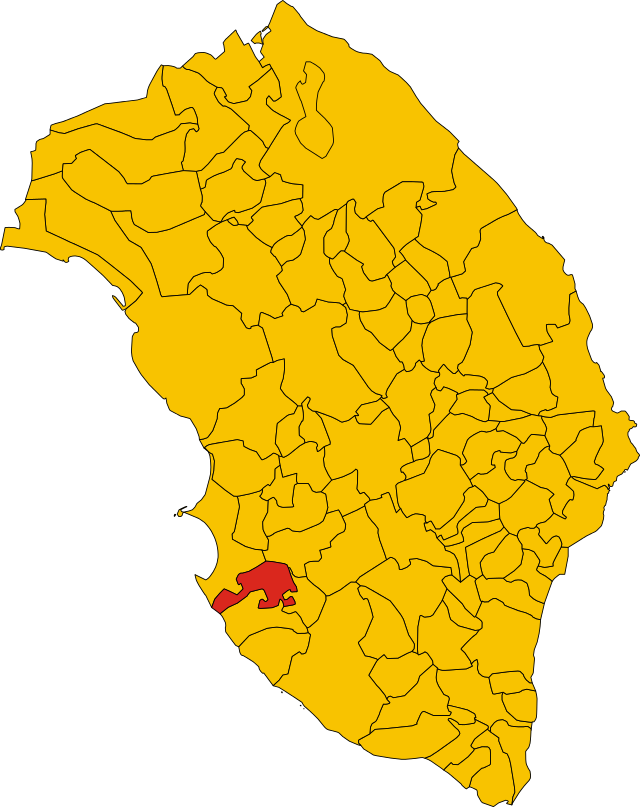}
\caption{Identification of Taviano in the Lecce's  province}
\end{center}
\end{figure}
\centering
\newpage
\begin{figure}
\begin{center}
\includegraphics[width=0.7\textwidth{}]{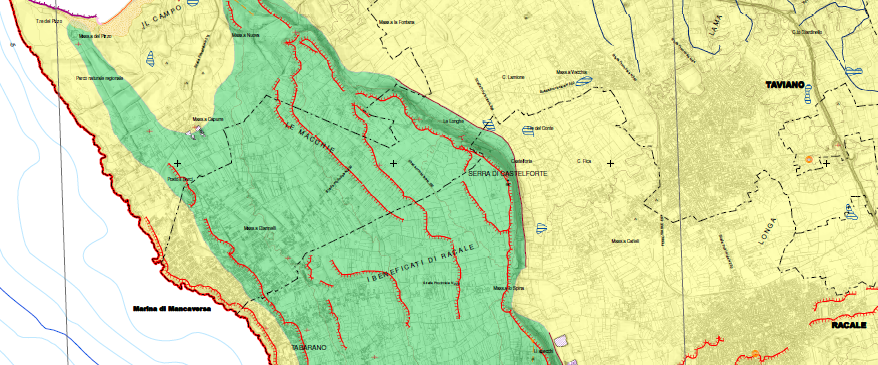}   \quad 
\includegraphics[width=0.2\textwidth]{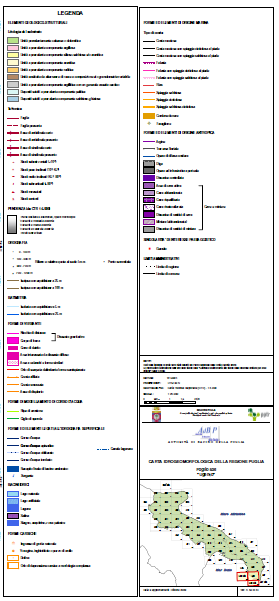}\\
\caption{Geological Map of Taviano - LE \cite{RegionePuglia:2015}} 
\end{center}
\end{figure}

\begin{table}
\tiny
\centering
\caption{Comparison of temperature on decadal scale} 
\begin{tabular}{|c|c|c|c|c|c|c|c|c|}
\hline
\bf Month & \bf Year &\bf Temperatures Min - Max&\bf Month & \bf Year &\bf  Temperatures Min - Max &\bf Month & \bf Year &\bf Temperatures Min - Max \cr

\hline
\bf Jannary &2014&2 - 16& \bf Febbrary &2014&3 - 19 &\bf March &2014&4 - 22\cr
& 2013 &0 - 17 & &2013 &-3 - 18 &&2013 &-1 - 20 \cr 
& 2012&-3 - 17 && 2012&-1 - 18 & & 2012&2 - 20 \cr
& 2011&-5 - 16 && 2011&0 - 17 & & 2011&-2 - 19 \cr
& 2010&1 - 19 && 2010&-1 - 19 & & 2010&-1 - 20 \cr
& 2009&0 - 16 && 2009&-2 - 16 & & 2009&1 - 21 \cr
& 2008&-2 - 18 && 2008&-2 - 19 & & 2008&2 - 20 \cr
& 2007&-2 - 21 && 2007&-2 - 18 & & 2007&0 - 20 \cr
& 2006&-6 - 16 && 2006&-4 - 20 & & 2006&-2 - 22 \cr
& 2005&-2 - 16 && 2005&-4 - 18 & & 2005&-2 - 21 \cr
& 2004&-3 - 17 && 2004&-2 - 18 & & 2004&-2 - 21 \cr
& 2003&1 - 19 && 2003&-3 - 16 & & 2003&0 - 21 \cr
& 2002&-3 - 16 && 2002&1 - 23 & & 2002&2 - 23 \cr 
& 2001&1 - 18 && 2001&-2 - 18 && 2001&3 - 29  \cr 
& 2000&-3 - 17 && 2000&-2 - 17 && 2000&-1 - 20  \cr
& 1999&-1 - 16 && 1999&-2 - 16 && 1999&2 - 20  \cr
& 1998&1 - 17 && 1998&1 - 19 && 1998&-3 - 20  \cr
& 1997&1 - 18 && 1997&-3 - 20 && 1997&1 - 21  \cr
& 1996&-2 - 16 && 1996&-3 - 17 && 1996&-2 - 19 \cr
& 1995&-1 - 19 && 1995&1 - 21 && 1995&2 - 19  \cr
& 1994&1 - 17 && 1994&1 - 17 && 1994&1 - 23 \cr
& 1993&1 - 16 && 1993&-3 - 15 && 1993&1 - 21 \cr
& 1992&-3 - 17 && 1992&-2 - 17 && 1992&0 - 18  \cr
& 1991&-2 - 17 && 1991&-5 - 18 && 1991&4 - 22  \cr
& 1990&-3 - 19 && 1990&0 - 19 && 1990&-2 - 24  \cr\hline
\bf April &2014&3 - 23& \bf May &2014&8 - 30&\bf June &2014&14 - 35\cr
& 2013&7 - 30 && 2013 &11 - 31 && 2013 &14 - 36 \cr 
& 2012&4 - 27 && 2012&7 - 29 & & 2012&15 - 36 \cr
& 2011&5 - 25 & & 2011&7 - 28 & & 2011&15 - 35 \cr
& 2010&5 - 24 & & 2010&8 - 28 & & 2010&11 - 35 \cr
& 2009&6 - 24 & & 2009&8 - 34 & & 2009&14 - 35 \cr
& 2008&4 - 24 & & 2008&7 - 34 & & 2008&15 - 34 \cr
& 2007&0 - 25 & & 2007&7 - 32 & & 2007&11 - 44 \cr
& 2006&2 - 26 & & 2006&6 - 36 & & 2006&10 - 40 \cr
& 2005&0 - 26 & & 2005&7 - 31 & & 2005&7 - 38 \cr
& 2004&3 - 23 & & 2004&3 - 28 & & 2004&11 - 36 \cr
& 2003&-2 - 24 & & 2003&9 - 33 & & 2003&16 - 37 \cr 
& 2002&1 - 22 & & 2002&7 - 26 & & 2002&11 - 37 \cr 
& 2001&2 - 25 & & 2001&8 - 34 & & 2001&9 - 33 \cr
& 2000&4 - 27 && 2000&11 - 31 && 2000&12 - 36  \cr
& 1999&3 - 29 && 1999&10 - 31 && 1999&11 - 35  \cr
& 1998&2 - 24 && 1998&12 - 29 && 1998&16 - 37  \cr
& 1997&-1 - 19 && 1997&6 - 31 && 1997&15 - 35  \cr
& 1996&5 - 24 && 1996&11 - 28 && 1996&15 - 33  \cr
& 1995&2 - 22 && 1995&7 - 30 && 1995&14 - 32  \cr
& 1994&4 - 23 && 1994&9 - 32 && 1994&11 - 35  \cr
& 1993&1 - 24 && 1993&7 - 31 && 1993&13 - 33  \cr
& 1992&4 - 27 && 1992&8 - 29 && 1992&13 - 33  \cr
& 1991&4 - 20 && 1991&5 - 26 && 1991&10 - 35  \cr
& 1990&5 - 23 && 1990&7 - 32 && 1990&11 - 35  \cr\hline
\bf July &2014&17 - 34& \bf August &2014&16 - 38&\bf September &2014&9 - 32\cr
&2013 &15 - 40 & & 2013 &19 - 38 && 2013 &14 - 33 \cr 
& 2012&18 - 40 & & 2012&16 - 38 & & 2012&10 - 36 \cr
& 2011&19 - 37 & & 2011&17 - 37 & & 2011&15 - 35 \cr
& 2010&16 - 37 & & 2010&2 - 36 & & 2010&11 - 30 \cr
& 2009&15 - 40 & & 2009&18 - 35 & & 2009&2 - 33 \cr
& 2008&18 - 35 & & 2008&17 - 39 & & 2008&10 - 37 \cr
& 2007&12 - 44 & & 2007&15 - 40 & & 2007&7 - 34 \cr
& 2006&15 - 36 & & 2006&10 - 39 & & 2006&15 - 32 \cr
& 2005&14 - 38 & & 2005&11 - 38 & & 2005&12 - 31 \cr
& 2004&14 - 40 & & 2004&11 - 37 & & 2004&7 - 31 \cr
& 2003&16 - 40 & & 2003&18 - 39 & & 2003&16 - 33 \cr 
& 2002&16 - 37 & & 2003&17 - 35 & & 2003&7 - 29 \cr 
& 2001&15 - 37 & & 2001&17 - 40 & & 2001&10 - 32 \cr
& 2000&14 - 41 & & 2000&16 - 39 && 2000&9 - 34  \cr
& 1999&16 - 34 && 1999&17 - 42 && 1999&16 - 31  \cr
& 1998&17 - 42 && 1998&15 - 38 && 1998&11 - 31  \cr
& 1997&15 - 39 && 1997&17 - 34 && 1997&11 - 32  \cr
& 1996&13 - 36 && 1996&16 - 34 && 1996&9 - 27  \cr
& 1995&17 - 37 && 1995&15 - 34 && 1995&11 - 29  \cr
& 1994&18 - 37 && 1994&17 - 39 && 1994&13 - 37 \cr
& 1993&10 - 40 && 1993&15 - 38 && 1993&11 - 32  \cr
& 1992&14 - 33 && 1992&17 - 37 && 1992&10 - 33  \cr
& 1991&16 - 36 && 1991&13 - 35 && 1991&13 - 31  \cr
& 1990&16 - 38 && 1990&15 - 37 && 1990&9 - 34  \cr\hline
\bf October &2014&7 - 27& \bf November &2014&5 - 23&\bf December &2014&-2 - 21\cr
&2013 &9 - 27 &&2013 &2 - 25 &&2013 &0 - 19 \cr 
& 2012&7 - 32 & & 2012&7 - 24 & & 2012&-2 - 18 \cr
& 2011&3 - 28 & & 2011&2 - 22 & & 2011&-2 - 19 \cr
& 2010&6 - 23 & & 2010&4 - 23 & & 2010&-3 - 20 \cr
& 2009&6 - 27 & & 2009&3 - 21 & & 2009&-1 - 21 \cr
& 2008&7 - 25 & & 2008&2 - 24 & & 2008&-2 - 18 \cr
& 2007&3 - 29 & & 2007&1 - 19 & & 2007&-3 - 17 \cr
& 2006&2 - 28 & & 2006&-1 - 23 & & 2006&-2 - 21 \cr
& 2005&2 - 26 & & 2005&6 - 22 & & 2005&-5 - 18 \cr
& 2004&10 - 28 & & 2004&-2 - 25 & & 2004&0 - 20 \cr
& 2003&8 - 29 & & 2003&5 - 23 & & 2003&-2 - 19 \cr
& 2002&5 - 25 & & 2003&3 - 22 & & 2003&-2 - 17 \cr
& 2001&8 - 31 & & 2001&1 - 24 & & 2001&-5 - 17 \cr
& 2000&7 - 28 && 2000&2 - 24 && 2000&-3 - 19  \cr
& 1999&9 - 29 && 1999&1 - 22 && 1999&2 - 19  \cr
& 1998&6 - 27 && 1998&1 - 23 && 1998&2 - 17  \cr
& 1997&3 - 27 && 1997&7 - 22 && 1997&2 - 17  \cr
& 1996&3 - 24 && 1996&5 - 22 && 1996&1 - 18  \cr
& 1995&4 - 27 && 1995&1 - 21 && 1995&3 - 18  \cr
& 1994&9 - 31 && 1994&3 - 23 && 1994&-2 - 19  \cr
& 1993&8 - 28 && 1993&3 - 22 && 1993&2 - 19  \cr
& 1992&10 - 27 && 1992&1 - 25 && 1992&0 - 20  \cr
& 1991&5 - 33 && 1991&4 - 22 && 1991&-3 - 15  \cr
& 1990&8 - 29 && 1990&2 - 27 && 1990&1 - 18  \cr 
\hline
\end{tabular}
\end{table}
\centering
\begin{table}
\centering \tiny
\caption{Day Time and Night Time Variation} 

\begin{tabular}{|c|c|c|c|c|c|c|c|c|c|c|c|c|}
\hline
\bf Year & \bf Jan &\bf Feb &\bf Mar & \bf Apr &\bf May&\bf Jun & \bf Jul &\bf Aug&\bf Sep&\bf Oct &\bf Nov&\bf Dec \cr
\hline
\bf 2014 & 6 & 6 & 9 & 9 & 9 & 10 & 9 & 10 & 8 & 9 & 8 & 8 \cr  \hline
\bf 2013 & 7 & 7 & 6 & 11 & 9 & 11 & 11 & 10 & 10 & 9 & 7 & 10 \cr  \hline
\bf 2012 & 8 & 6 & 9 & 7 & 10 & 11 & 12 & 11 & 11 & 10 & 8 & 6 \cr  \hline
\bf 2011 & 8 & 8 & 7 & 10 & 9 & 9 & 10 & 11 & 10 & 8 & 9 & 8 \cr  \hline
\bf 2010 & 5 & 7 & 10 & 10 & 9 & 11 & 10 & 12 & 9 & 7 & 7 & 7 \cr  \hline
\bf 2009 & 6 & 7 & 8 & 8 & 11 & 9 & 11 & 11 & 10 & 8 & 7 & 6 \cr  \hline
\bf 2008 & 8 & 9 & 10 & 10 & 12 & 10 & 10 & 12 & 9 & 9 & 8 & 6 \cr  \hline
\bf 2007 & 11 & 8 & 10 & 13 & 13 & 14 & 15 & 14 & 12 & 9 & 7 & 7 \cr  \hline
\bf 2006 & 10 & 9 & 9 & 12 & 13 & 14 & 12 & 13 & 12 & 10 & 13 & 11 \cr  \hline
\bf 2005 & 9 & 10 & 11 & 11 & 12 & 12 & 13 & 13 & 10 & 10 & 11 & 8 \cr  \hline
\bf 2004 & 8 & 9 & 9 & 10 & 13 & 13 & 14 & 13 & 12 & 10 & 9& 7 \cr  \hline
\bf 2003 & 8 & 9 & 11 & 10 & 12 & 14 & 13 & 14 & 11 & 9 & 7 & 7 \cr  \hline
\bf 2002 & 9 & 10 & 8 & 10 & 10 & 13 & 11 & 11 & 9 & 9 & 8 & 7 \cr  \hline
\bf 2001 & 8 & 10 & 11 & 10 & 11 & 12 & 13 & 13 & 11 & 11 & 8 & 7  \cr  \hline
\bf 2000 & 9 & 8 & 9 & 10 & 11 & 14 & 15 & 14 & 11 & 9 & 8 & 9  \cr  \hline
\bf 1999 & 8 & 7 & 8 & 9 & 11 & 10 & 11 & 13 & 9 & 9 & 7 & 6  \cr  \hline
\bf 1998 & 7 & 9 & 8 & 10 & 9 & 11 & 12 & 12 & 9 & 8 & 7 & 5  \cr  \hline
\bf 1997 & 6 & 10 & 9 & 8 & 10 & 10 & 12 & 9 & 9 & 8 & 6 & 6  \cr  \hline
\bf 1996 & 6 & 6 & 7 & 8 & 9 & 9 & 10 & 11 & 8 & 9 & 7 & 6  \cr  \hline
\bf 1995 & 7 & 8 & 8 & 9 & 9 & 8 & 10 & 8 & 9 & 10 & 7 & 5  \cr  \hline
\bf 1994 & 7 & 7 & 11 & 9 & 10 & 10 & 11 & 13 & 12 & 8 & 8 & 8  \cr  \hline
\bf 1993 & 8 & 6 & 9 & 14 & 19 & 23 &  25 & 27 & 22 & 19 & 13 & 12  \cr  \hline
\bf 1992 & 10 & 10 & 10 & 9 & 12 & 11 & 11 & 13 & 12 & 7 & 10 & 9  \cr  \hline
\bf 1991 & 9 & 8 & 8 & 9 & 11 & 14 & 12 & 12 & 11 & 8 & 9 & 8  \cr  \hline
\bf 1990 & 10 &  10 & 14 & 10 & 11 & 12 &    13 & 13 & 11 & 9 & 8 & 7  \cr  \hline

\hline
\bf Average 1990-2014 & 7,94 & 8,20  & 9,13 & 9,89 & 11,19 & 11,80 & 12,23 & 12,55 & 10,62 & 9,42 & 8,31 & 7,51 \cr  \hline

\end{tabular}
\end{table}


\begin{table}
\tiny
\centering
\caption{Rain/Year [mm]}

\begin{tabular}{|c|c|c|c|c|c|c|c|c|c|c|c|c|c|c|c|}
\hline
\bf Year & & Jan & Feb & Mar & Apr & May & Jun & Jul & Ago & Sep & Oct & Nov &  Dec & \bf Tot & \bf \%Year \\
\hline
&  &  &  & &  & & & & & & & & & &          \cr
\hline
2014 &   {\bf 58} &      55,00 &      88,00 &      67,00 &     100,00 &      60,00 &      55,00 &      30,00 &      10,00 &     100,00 &      45,00 &      50,00 &      71,00 & {\bf 731,00} & {\bf 4,57\%} \cr
\hline
      2013 &   {\bf 46} &      60,00 &      90,00 &      55,00 &      20,00 &      25,00 &      20,00 &      55,00 &      45,00 &      50,00 &      30,00 &      95,00 &      20,00 & {\bf 565,00} & {\bf 3,53\%} \cr
\hline
      2012 &   {\bf 24} &      20,00 &      60,00 &      15,00 &      60,00 &      20,00 &       5,00 &       5,00 &       5,00 &      45,00 &      75,00 &      35,00 &      45,00 & {\bf 390,00} & {\bf 2,44\%} \cr
\hline
      2011 &   {\bf 33} &      25,00 &      45,00 &      45,00 &      35,00 &      55,00 &      50,00 &       0,00 &      10,00 &      35,00 &      20,00 &      35,00 &      15,00 & {\bf 370,00} & {\bf 2,31\%} \cr
\hline
      2010 &   {\bf 31} &      55,00 &      55,00 &      25,00 &      40,00 &      15,00 &      30,00 &      25,00 &       0,00 &      50,00 &      65,00 &      40,00 &      20,00 & {\bf 420,00} & {\bf 2,62\%} \cr
\hline
      2009 &   {\bf 98} &     279,00 &      36,50 &      95,70 &      72,60 &      40,40 &     180,00 &      10,00 &      68,10 &      55,00 &      40,00 &      40,00 &      10,00 & {\bf 927,30} & {\bf 5,79\%} \cr
\hline
      2008 &   {\bf 59} &      58,20 &      32,70 &      87,90 &      17,00 &      49,50 &       2,00 &       3,10 &     139,20 &      43,70 &      67,50 &     246,40 &      44,20 & {\bf 791,40} & {\bf 4,95\%} \cr
\hline
      2007 &   {\bf 71} &      13,10 &      42,60 &     141,60 &      38,60 &      58,80 &      31,60 &       0,50 &      45,00 &     183,50 &     138,90 &      57,10 &      27,40 & {\bf 778,70} & {\bf 4,87\%} \cr
\hline
      2006 &   {\bf 46} &      25,70 &      80,90 &      46,30 &      34,90 &      74,10 &      72,20 &      72,20 &      10,00 &      80,70 &      10,00 &      42,70 &       0,50 & {\bf 550,20} & {\bf 3,44\%} \cr
\hline
      2005 &   {\bf 46} &      65,40 &      59,80 &      57,10 &      46,00 &      29,70 &       2,00 &      25,20 &     101,00 &      67,30 &      34,60 &      34,60 &      31,40 & {\bf 554,10} & {\bf 3,46\%} \cr
\hline
      2004 &   {\bf 97} &      70,70 &      32,30 &     168,70 &      44,60 &     119,10 &       3,10 &      98,10 &      23,10 &      94,70 &     343,00 &     133,60 &      35,60 & {\bf 1166,60} & {\bf 7,29\%} \cr
\hline
      2003 &   {\bf 62} &     110,40 &      15,90 &       1,00 &      30,80 &      68,70 &      30,20 &      13,60 &     105,70 &     122,90 &     105,60 &     112,10 &      32,00 & {\bf 748,90} & {\bf 4,68\%} \cr
\hline
      2002 &   {\bf 53} &      32,00 &       4,60 &      17,80 &     101,40 &      13,10 &      26,80 &      67,80 &      51,00 &     127,90 &      63,90 &      25,00 &     104,60 & {\bf 635,90} & {\bf 3,97\%} \cr
\hline
      2001 &   {\bf 31} &     116,10 &       9,10 &      39,70 &      33,30 &      30,80 &      15,70 &       5,10 &      15,00 &       3,50 &      27,00 &      33,70 &      48,40 & {\bf 377,40} & {\bf 2,36\%} \cr
\hline
      2000 &   {\bf 50} &       5,00 &      45,00 &      15,00 &      30,00 &      45,00 &      26,10 &       5,10 &       0,00 &      22,80 &     214,70 &     109,70 &      85,80 & {\bf 604,20} & {\bf 3,78\%} \cr
\hline
      1999 &   {\bf 44} &      25,00 &      45,00 &      40,00 &      78,00 &       0,00 &      50,00 &      30,00 &      20,00 &      95,00 &      20,00 &      70,00 &      50,00 & {\bf 523,00} & {\bf 3,27\%} \\
\hline
      1998 &   {\bf 36} &      25,00 &      20,00 &      35,00 &      15,00 &      45,00 &      20,00 &      15,00 &      45,00 &      50,00 &      65,00 &      45,00 &      50,00 & {\bf 430,00} & {\bf 2,69\%} \cr
\hline
      1997 &   {\bf 28} &      40,00 &      15,00 &      25,00 &      30,00 &      10,00 &      15,00 &       0,00 &      50,00 &      35,00 &      50,00 &      30,00 &      30,00 & {\bf 330,00} & {\bf 2,06\%} \cr
\hline
      1996 &   {\bf 66} &      85,00 &      55,00 &      75,00 &      85,00 &      40,00 &      20,00 &       0,00 &     120,00 &      90,00 &     115,00 &      45,00 &      65,00 & {\bf 795,00} & {\bf 4,97\%} \cr
\hline
      1995 &   {\bf 55} &      60,00 &      35,00 &      60,00 &      35,00 &      15,00 &       5,00 &      60,00 &     185,00 &      45,00 &       5,00 &     100,00 &      60,00 & {\bf 665,00} & {\bf 4,16\%} \cr
\hline
      1994 &   {\bf 41} &     100,00 &      65,00 &      10,00 &      65,00 &      25,00 &      55,00 &      60,00 &      10,00 &      50,00 &      10,00 &      30,00 &      15,00 & {\bf 495,00} & {\bf 3,09\%} \cr
\hline
      1993 &   {\bf 55} &      35,00 &      60,00 &      85,00 &      55,00 &      80,00 &      30,00 &      18,00 &      10,00 &      65,00 &      60,00 &     110,00 &      55,00 & {\bf 663,00} & {\bf 4,14\%} \cr
\hline
      1992 &   {\bf 62} &      35,00 &      20,00 &      50,00 &     110,00 &      30,00 &     130,00 &      68,00 &      10,00 &      80,00 &      95,00 &      40,00 &      75,00 & {\bf 743,00} & {\bf 4,64\%} \cr
\hline
      1991 &   {\bf 90} &      40,00 &      75,00 &      70,00 &     137,00 &      85,00 &      32,00 &      55,00 &      40,00 &     145,00 &     155,00 &     135,00 &     110,00 & {\bf 1079,00} & {\bf 6,74\%} \cr
\hline
      1990 &   {\bf 56} &      15,00 &      10,00 &      35,00 &      80,00 &      80,00 &      20,00 &       5,00 &      50,00 &      25,00 &      55,00 &     120,00 &     175,00 & {\bf 670,00} & {\bf 4,19\%} \cr
\hline
\bf Average 1990-2014 & \bf 53,56 & \bf 58,02 & \bf 43,90 & \bf 54,51 & \bf 55,77 & \bf 44,57 & \bf 37,07 & \bf 29,07 & \bf 46,72 & \bf 70,48 & \bf 76,41 & \bf 72,60 & \bf 51,04 & \bf 16003,70 &  \cr
\hline
\bf &&&&&&&&&&&&&& \bf Year Average & \bf 640,15 \\
\hline
\end{tabular}
\end{table}
\begin{table}
\caption{Wind Average[1990-2014]/Year on Km/h}
\tiny
\begin{tabular}{|c|c|c|c|c|c|c|c|c|c|c|c|c|c|c|c|} 
\hline
\multicolumn{ 14}{|c}{} &            &            \\
\hline
\multicolumn{ 1}{|c|}{Day} &        Jan &        Feb &        Mar &        Apr &        May &        Jun &        Jul &        Ago &        Sep &        Oct &        Nov &        Dec &\bf Average &  \bf Min &  \bf Max  \\
\hline
\multicolumn{ 1}{|c|}{} &            &            &            &            &            &            &            &            &            &            &            &            &            &            &            \\

         1 &      24,91 &      24,21 &      32,44 &      24,08 &      19,48 &      22,86 &      23,44 &      25,35 &      25,42 &      24,00 &      22,13 &      25,56 & \bf 24,49 &      19,48 &      32,44 \cr 
         2 &      21,13 &      26,68 &      28,33 &      24,28 &      20,91 &      22,52 &      24,92 &      21,87 &      22,65 &      20,71 &      24,61 &      23,60 & \bf 23,52 &      20,71 &      28,33 \cr
         3 &      24,13 &      26,32 &      27,71 &      25,84 &      28,44 &      23,64 &      20,83 &      21,92 &      22,27 &      25,83 &      25,58 &      23,84 & \bf 24,70 &      20,83 &      28,44 \cr
         4 &      25,54 &      28,12 &      30,22 &      26,80 &      26,64 &      25,26 &      21,96 &      23,08 &      24,23 &      24,21 &      26,00 &      22,32 & \bf 25,36 &      21,96 &      30,22 \cr
         5 &      25,08 &      26,48 &      30,60 &      28,63 &      24,24 &      22,13 &      23,35 &      27,22 &      28,79 &      21,71 &      26,21 &      23,13 & \bf 25,63 &      21,71 &      30,60 \cr
         6 &      26,68 &      29,36 &      29,88 &      29,08 &      21,71 &      24,68 &      23,83 &      25,00 &      24,17 &      19,40 &      27,75 &      25,04 & \bf 25,55 &      19,40 &      29,88 \cr
         7 &      24,21 &      27,88 &      24,44 &      26,44 &      25,17 &      24,87 &      24,84 &      23,57 &      20,59 &      25,52 &      27,13 &      23,70 & \bf 24,86 &      20,59 &      27,88 \cr
         8 &      22,84 &      26,36 &      29,68 &      27,13 &      23,42 &      24,26 &      29,12 &      22,95 &      23,87 &      25,50 &      21,96 &      23,70 & \bf 25,07 &      21,96 &      29,68 \cr
         9 &      22,91 &      24,48 &      27,12 &      31,40 &      22,22 &      25,09 &      25,21 &      24,35 &      25,17 &      23,08 &      22,41 &      22,20 & \bf 24,64 &      22,20 &      31,40 \cr
        10 &      22,65 &      29,16 &      27,16 &      31,40 &      23,48 &      24,46 &      22,87 &      24,13 &      23,39 &      23,28 &      23,52 &      26,32 & \bf 25,15 &      22,65 &      31,40 \cr
        11 &      20,43 &      27,48 &      24,72 &      28,60 &      21,29 &      24,05 &      27,09 &      24,13 &      22,30 &      23,67 &      23,91 &      21,17 & \bf 24,07 &      20,43 &      28,60 \cr
        12 &      19,83 &      26,68 &      22,48 &      26,80 &      22,30 &      27,15 &      24,61 &      27,74 &      23,22 &      23,13 &      25,96 &      22,42 & \bf 24,36 &      19,83 &      27,74 \cr
        13 &      22,83 &      25,72 &      25,36 &      28,04 &      24,63 &      23,00 &      24,33 &      24,13 &      25,75 &      25,36 &      29,00 &      21,10 & \bf 24,94 &      21,10 &      29,00 \cr
        14 &      22,29 &      27,92 &      21,92 &      24,36 &      22,96 &      23,04 &      24,08 &      20,00 &      23,21 &      24,08 &      24,35 &      24,30 & \bf 23,54 &      20,00 &      27,92 \cr
        15 &      26,75 &      24,64 &      21,44 &      26,72 &      25,08 &      24,70 &      27,75 &      22,74 &      23,00 &      22,12 &      22,55 &      26,12 & \bf 24,47 &      21,44 &      27,75 \cr
        16 &      22,04 &      28,72 &      25,96 &      27,54 &      23,48 &      23,30 &      29,21 &      21,04 &      22,71 &      19,87 &      26,00 &      24,24 & \bf 24,51 &      19,87 &      29,21 \cr
        17 &      23,91 &      29,76 &      26,84 &      28,84 &      21,65 &      25,63 &      23,96 &      22,88 &      21,83 &      20,84 &      24,78 &      26,92 & \bf 24,82 &      20,84 &      29,76 \cr
        18 &      22,17 &      29,60 &      24,08 &      27,12 &      26,30 &      23,52 &      25,91 &      22,21 &      20,42 &      19,71 &      22,78 &      22,48 & \bf 23,86 &      19,71 &      29,60 \cr
        19 &      25,96 &      31,00 &      23,88 &      29,12 &      24,65 &      25,05 &      27,50 &      21,65 &      21,04 &      22,17 &      25,13 &      21,08 & \bf 24,85 &      21,04 &      31,00 \cr
        20 &      24,32 &      30,20 &      23,56 &      24,72 &      24,87 &      25,57 &      27,09 &      22,00 &      25,50 &      21,08 &      24,43 &      22,87 & \bf 24,68 &      21,08 &      30,20 \cr
        21 &      20,60 &      26,28 &      24,92 &      25,72 &      24,18 &      24,09 &      27,08 &      21,48 &      23,08 &      24,67 &      25,22 &      20,09 & \bf 23,95 &      20,09 &      27,08 \cr
        22 &      24,76 &      27,42 &      27,04 &      24,50 &      26,22 &      24,48 &      27,67 &      23,52 &      22,29 &      25,73 &      25,32 &      26,39 & \bf 25,44 &      22,29 &      27,67 \cr
        23 &      21,80 &      26,88 &      30,08 &      24,00 &      21,52 &      23,61 &      23,42 &      22,83 &      24,40 &      20,91 &      23,23 &      25,78 & \bf 24,04 &      20,91 &      30,08 \cr
        24 &      25,96 &      29,12 &      27,08 &      22,28 &      26,50 &      22,79 &      21,95 &      24,54 &      24,58 &      21,13 &      26,17 &      19,96 & \bf 24,34 &      19,96 &      29,12 \cr
        25 &      23,74 &      23,16 &      28,64 &      22,00 &      25,00 &      26,21 &      25,14 &      21,96 &      23,09 &      21,68 &      25,13 &      25,24 & \bf 24,25 &      21,68 &      28,64 \cr
        26 &      23,04 &      24,68 &      27,60 &      21,48 &      20,55 &      24,57 &      25,90 &      22,50 &      25,56 &      20,91 &      24,26 &      30,14 & \bf 24,27 &      20,55 &      30,14 \cr
        27 &      24,46 &      23,50 &      29,64 &      25,75 &      23,70 &      23,43 &      26,14 &      23,91 &      24,33 &      23,78 &      25,35 &      30,13 & \bf 25,34 &      23,43 &      30,13 \cr
        28 &      23,91 &      31,92 &      26,40 &      20,32 &      23,58 &      23,33 &      23,46 &      24,55 &      22,38 &      24,65 &      28,30 &      25,32 & \bf 24,84 &      20,32 &      31,92 \cr
        29 &      25,79 &      29,57 &      24,52 &      25,12 &      24,68 &      21,44 &      23,95 &      26,14 &      20,72 &      25,78 &      22,27 &      21,96 & \bf 24,33 &      20,72 &      29,57 \cr
        30 &      26,75 &            &      30,29 &      23,63 &      22,21 &      23,48 &      26,29 &      22,68 &      20,96 &      25,17 &      25,23 &      20,48 & \bf 24,29 &      20,48 &      30,29 \cr
        31 &      25,91 &            &      25,76 &            &      22,75 &            &      24,61 &      24,83 &            &      26,58 &            &      26,78 & \bf 25,32 &      22,75 &      26,78 \cr
\hline
{\bf Average} & {\bf 23,79} & {\bf 27,35} & {\bf 26,77} & {\bf 26,06} & {\bf 23,67} & {\bf 24,07} & {\bf 25,08} & {\bf 23,45} & {\bf 23,36} & {\bf 23,10} & {\bf 24,89} & {\bf 24,01} & {\bf Media} & {\bf 19,40} & {\bf 32,44} \\
\hline
 {\bf Min} &      19,83 &      23,16 &      21,44 &      20,32 &      19,48 &      21,44 &      20,83 &      20,00 &      20,42 &      19,40 &      21,96 &      19,96 & {\bf 23,52} &            &            \\
\hline
 {\bf Max} &      26,75 &      31,92 &      32,44 &      31,40 &      28,44 &      27,15 &      29,21 &      27,74 &      28,79 &      26,58 &      29,00 &      30,14 & {\bf 25,63} &            &            \\
\hline
\end{tabular}  
\end{table}
\centering
\begin{table}
\centering \tiny
\caption{Trend of Temperature of Taviano in Jan-Feb-Mar and Jun-Jul-Aug} 
\begin{tabular}{|c|c|c|c|c|c|c|c|c|}
\hline
\bf Year & \bf Jan & \bf Feb &\bf Mar&\bf Jun & \bf Jul &\bf Aug \cr
\hline
\bf 2014 & 11,40  & 12,50  & 11,74  &24,58 &25,74  &26,52 \cr  \hline
\bf 2013 &9,55  &8,82  &12,40  &23,87 &26,55 &27,87 \cr  \hline
\bf 2012 &7,22  &4,43  &12,03  &25,32 &28,16 &27,37 \cr  \hline
\bf 2011 &8,42  &9,05  &11,11  &24,52 &26,26 &26,87 \cr  \hline
\bf 2010 &9,85  &10,64  &10,17  &22,52 &25,83 &25,23 \cr  \hline
\bf 2009 &9,77  &7,70  &10,65  &23,25 &25,66 &26,39 \cr  \hline
\bf 2008 &9,54  &8,92  &11,93  &24,15 &26,69 &26,87 \cr  \hline
\bf 2007 &10,35  &10,63  &11,62  &24,15 &26,46 &27,19 \cr  \hline
\bf 2006 &6,59  &9,23  &10,79  &22,08 &24,91 &24,52 \cr  \hline
\bf 2005 &7,95  &7,28  &10,25  &22,75 &26,70 &23,97 \cr  \hline
\bf 2004 &7,74  &8,76  &10,41  &22,62 &25,23 &25,14 \cr  \hline
\bf 2003 &10,58  &5,86  &9,70  &26,16 &27,37 &27,87 \cr  \hline
\bf 2002 &6,40  &11,80  &12,50  &23,60 &26,90 &25,90 \cr  \hline
\bf 2001 &11,25  &9,83  &14,80  &22,40 &25,91 &26,62 \cr  \hline
\bf 2000 &6,82  &8,44  &10,65  &24,03 &24,98 &26,55 \cr  \hline
\bf 1999 &8,24  &8,09  &11,44  &23,83 &25,44 &27,10 \cr  \hline
\bf 1998 &9,44  &10,77  &9,44  &24,82 &27,57 &27,35 \cr  \hline
\bf 1997 &10,31 &9,91  &11,28  &24,71 &25,82 &25,20 \cr  \hline
\bf 1996 &9,58 &8,39 &9,71  &24,46  &25,72 &24,95 \cr  \hline
\bf 1995 &8,62  &11,31  &10,53  &24,41 &27,44 &24,52 \cr  \hline
\bf 1994 &10,05  &9,73  &11,87  &22,02 &26,87 &27,65 \cr  \hline
\bf 1993 &8,07  &6,50  &9,05  &22,95 &24,84 &26,45 \cr  \hline
\bf 1992 &7,76  &7,47  &10,27  &22,02 &24,16 &27,24 \cr  \hline
\bf 1991 &7,89  &8,57  &12,82  &22,43 &24,97 &25,31  \cr  \hline
\bf 1990 &7,47  &9,62  &12,54  &23,26 &26,00 &25,11  \cr 
\hline
\bf Average 1990-2014 &8,83 &8,97 &11,19 &23,64 &26,09 &26,23  \cr  \hline

\end{tabular}
\end{table}
\newpage \clearpage

\begin{figure}
\begin{center}
\includegraphics[width=1.3\textwidth{}]{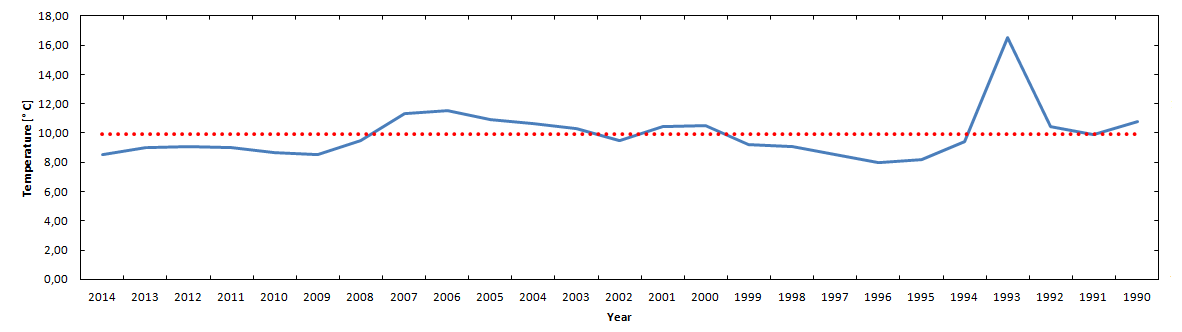}
\caption{Day time and night time variations (solid line) and Average Day time and Night time variations 1990-2014 (dotted line)}
\end{center}
\end{figure}
\begin{figure}
\begin{center}
\includegraphics[width=1.3\textwidth{}]{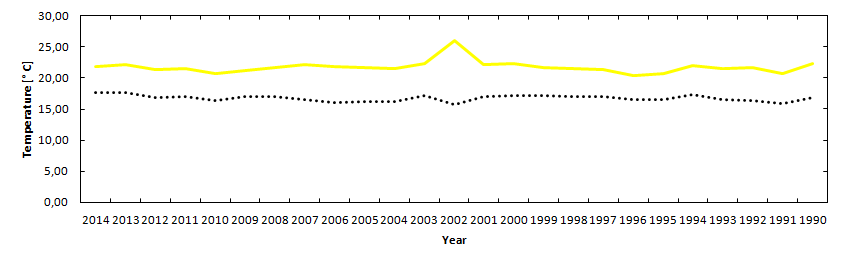}
\caption{Maximum Temperature (solid line) and Average Temperature Variations 1990-2014 (dotted line)}
\end{center}
\end{figure}
\begin{figure}
\begin{center}
\includegraphics[width=1.3\textwidth{}]{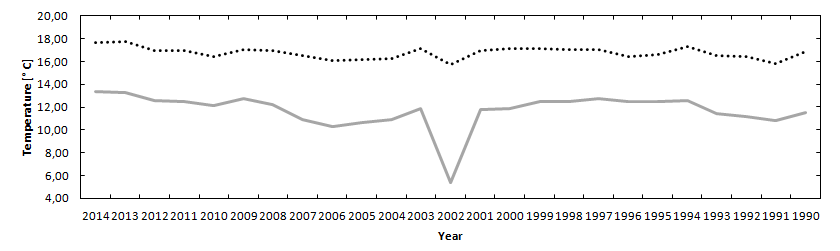}
\caption{Minimum Temperature (solid line) and Average Temperature Variations 1990-2014 (dotted line)}
\end{center}
\end{figure}
\begin{figure}
\begin{center}
\includegraphics[width=1.3\textwidth{}]{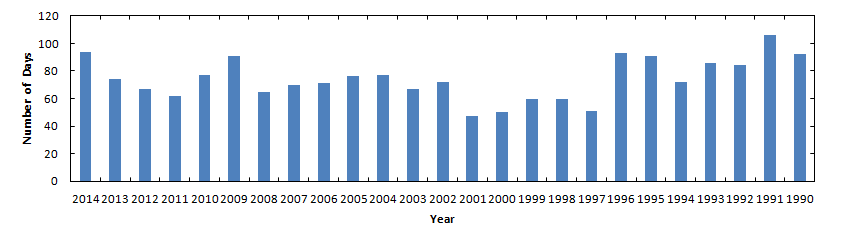}
\caption{Total Days of Rain by Year}
\end{center}
\end{figure}
\begin{figure}
\begin{center}
\includegraphics[width=1.3\textwidth{}]{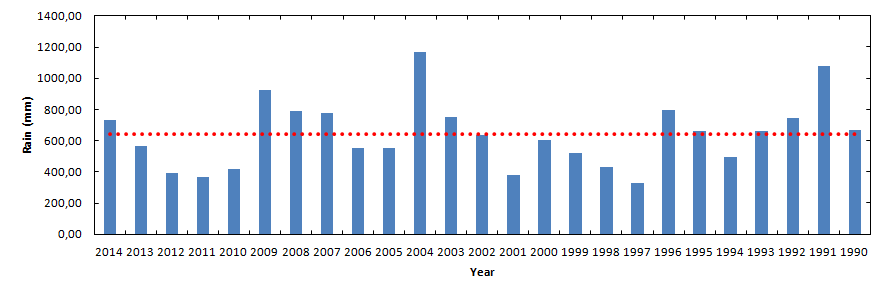}
\caption{Total rain [mm] by year - (Dotted line is Average from 1990 to 2014)}
\end{center}
\end{figure}
\begin{figure}
\begin{center}
\includegraphics[width=1.3\textwidth{}]{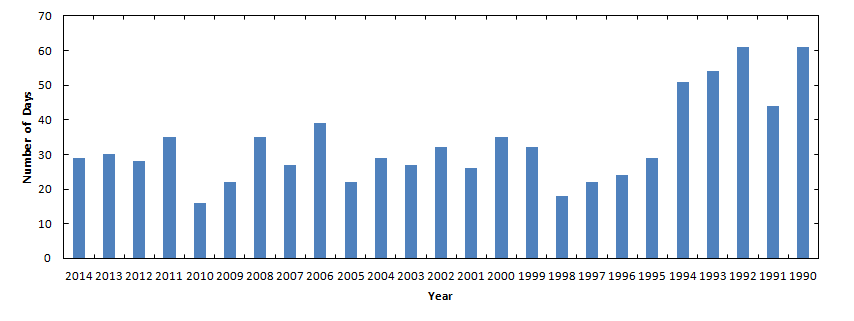}
\caption{Total day of Fog by year}
\end{center}
\end{figure}
\begin{figure}
\begin{center}
\includegraphics[width=1.3\textwidth{}]{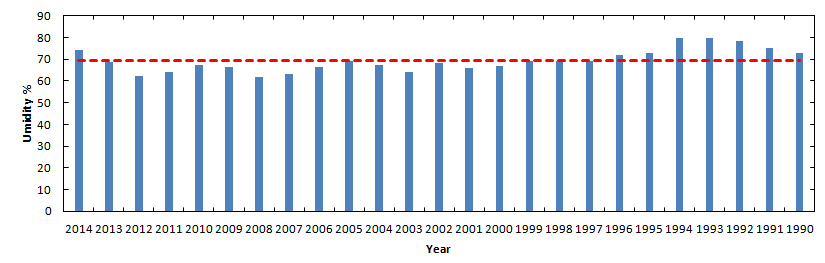}
\caption{Umidity (solid line) and Average Umidity Variations 1990-2014 (dotted line) }
\end{center}
\end{figure}
\begin{figure}
\begin{center}
\includegraphics[width=1.3\textwidth{}]{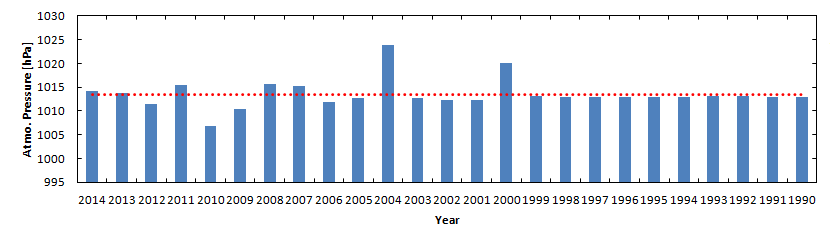}
\caption{Pressure (solid line) and Average Pressure Variations 1990-2014 (dotted line) }
\end{center}
\end{figure}
\begin{figure}
\begin{center}
\includegraphics[width=1.3\textwidth{}]{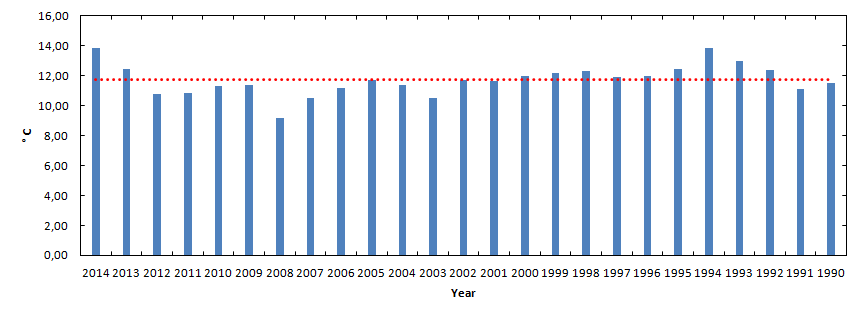}
\caption{Dew Point (solid line) and Average Dew Point Variations 1990-2014 (dotted line) }
\end{center}
\end{figure}
\begin{figure}
\begin{center}
\includegraphics[width=1.3\textwidth{}]{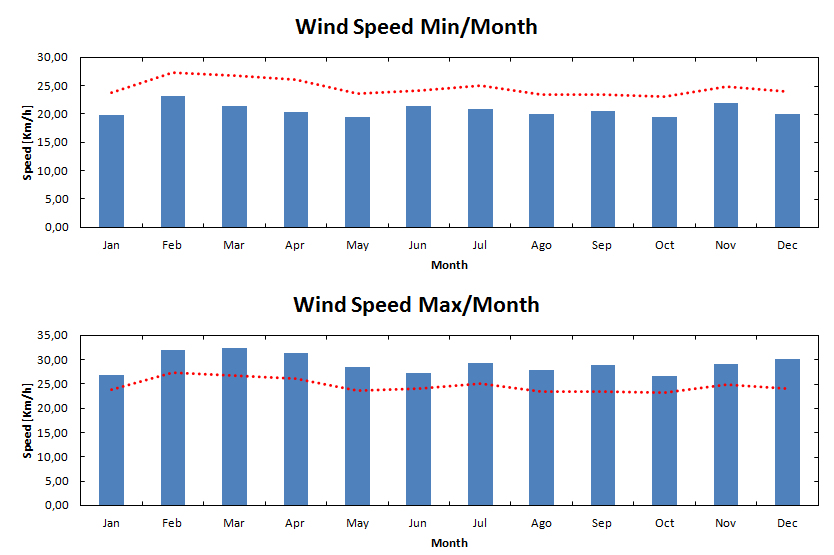}
\caption{Wind Speed (solid line) and Average Wind Speed Variations 1990-2014 (dotted line) }
\end{center}
\end{figure}
\clearpage
\begin{figure}
\includegraphics[width=1.3\textwidth{}]{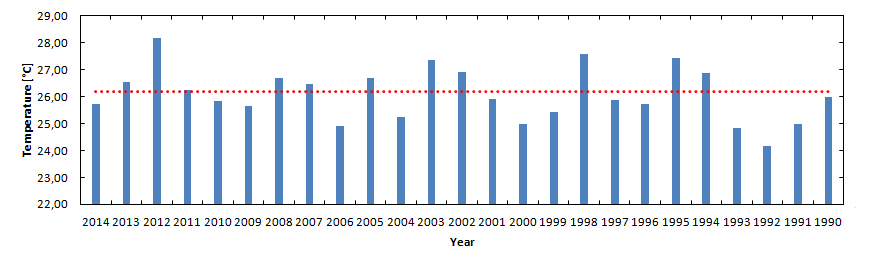}
\caption{July Temperature (solid line) and Average August Temperature 1990-2014 (dotted line) }
\end{figure}
\begin{figure}
\includegraphics[width=1.3\textwidth{}]{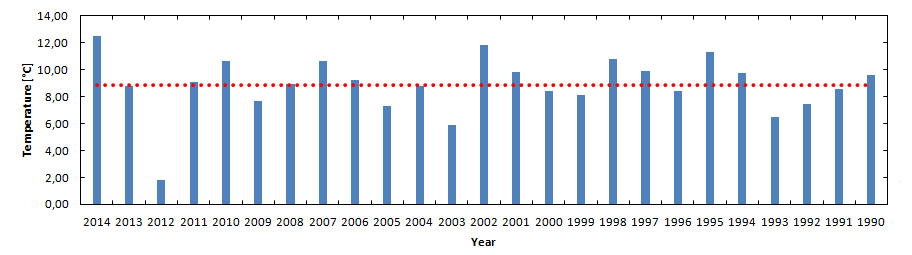}
\caption{February Temperature (solid line) and Average February Temperature 1990-2014 (dotted line)}
\end{figure}

\centering
\begin{figure}
\begin{center}
\includegraphics[width=1.3\textwidth{}]{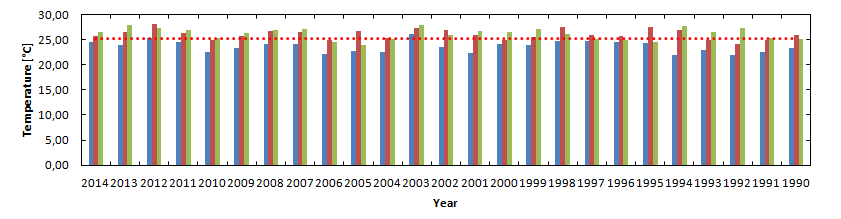}
\caption{Summer Temperature (solid line) and Average Temperature Variations 1990-2014 (dotted line). Jun: blu, Jul: red Aug: green}
\end{center}
\end{figure}
\begin{figure}
\begin{center}
\includegraphics[width=1.3\textwidth{}]{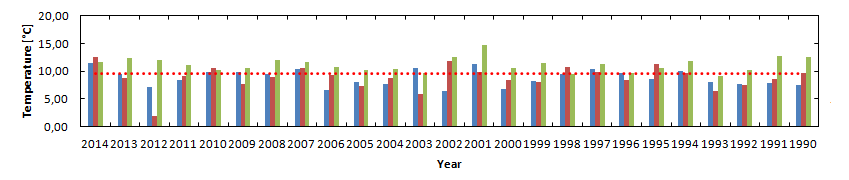}
\caption{Winner Temperature (solid line) and Average Temperature Variations 1990-2014 (dotted line) Jan: blu, Feb: red Mar: green}
\end{center}
\end{figure}
\begin{figure}
\caption{Total Rain (mm) by Month}
\includegraphics[width=0.55\textwidth{}]{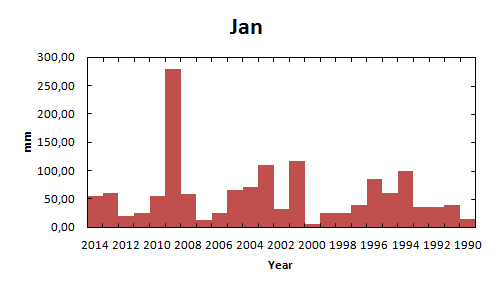} \quad 
\includegraphics[width=0.55\textwidth{}]{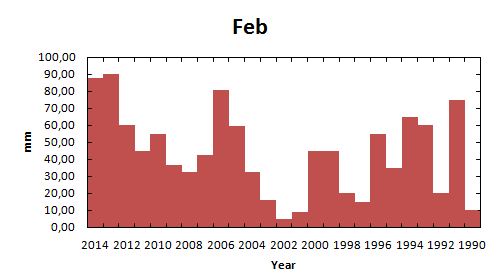} \quad
\includegraphics[width=0.55\textwidth{}]{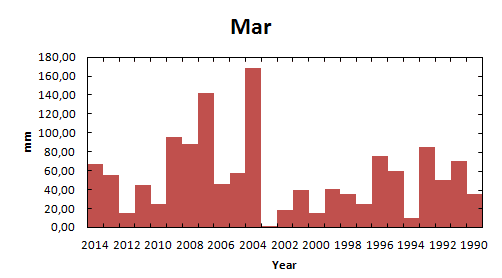} \quad
\includegraphics[width=0.60\textwidth{}]{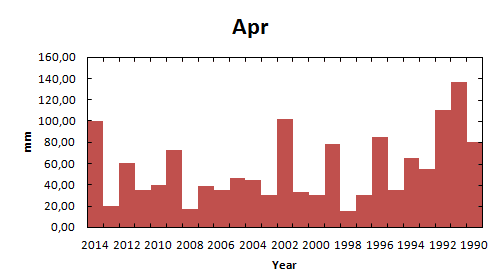} \quad
\includegraphics[width=0.55\textwidth{}]{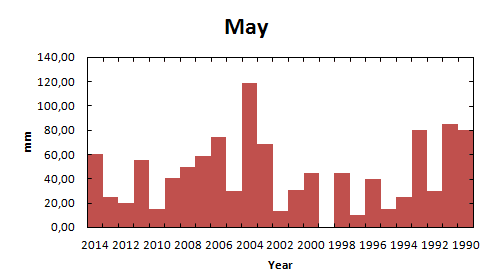} \quad 
\includegraphics[width=0.55\textwidth{}]{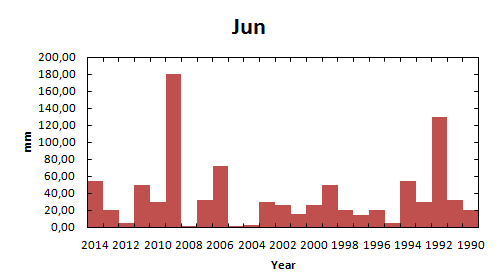} \quad
\includegraphics[width=0.55\textwidth{}]{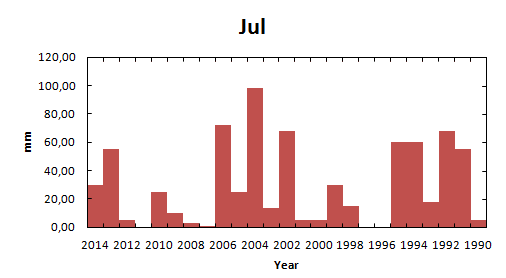} \quad
\includegraphics[width=0.55\textwidth{}]{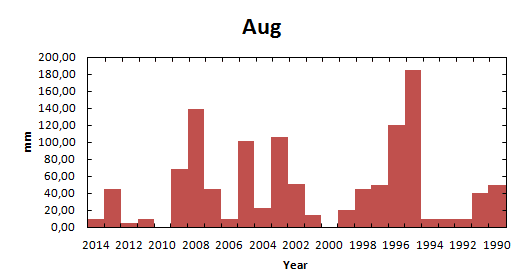} \quad
\includegraphics[width=0.55\textwidth{}]{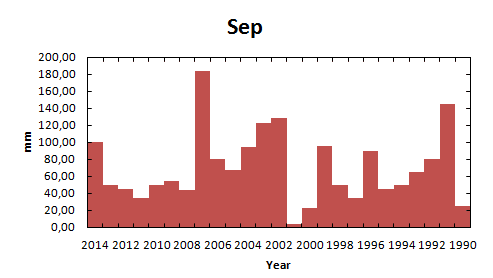} \quad 
\includegraphics[width=0.55\textwidth{}]{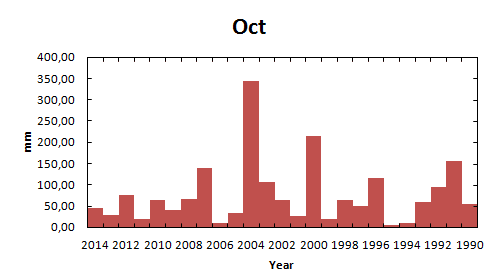} \quad
\includegraphics[width=0.55\textwidth{}]{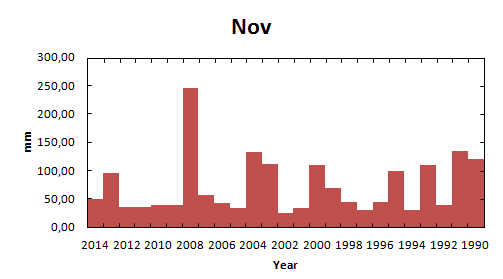} \quad
\includegraphics[width=0.55\textwidth{}]{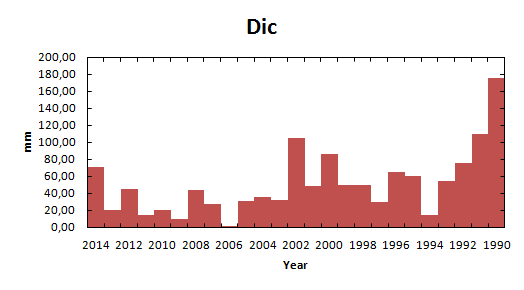} \\
\end{figure}
\begin{figure}
\caption{Solar Irradiation Map \cite{UE:2013}}
\includegraphics[width=1\textwidth{}]{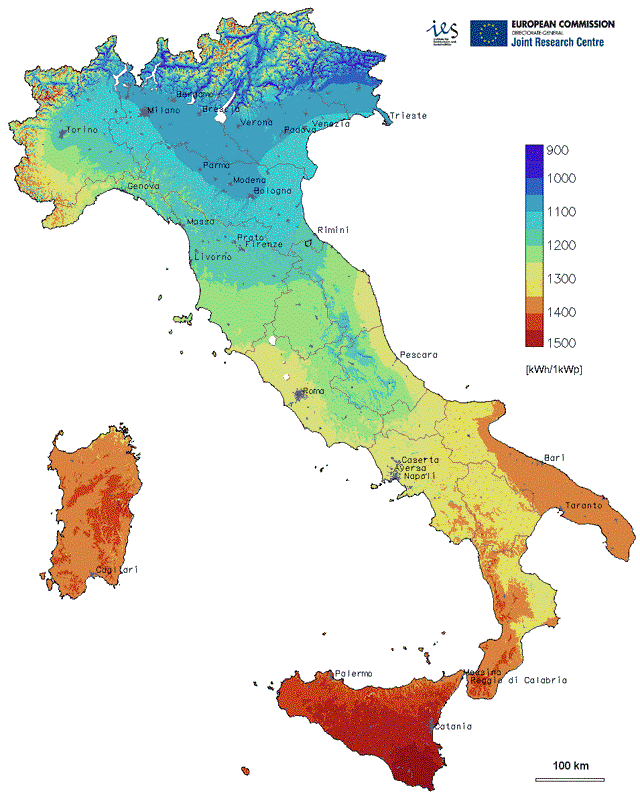}
\end{figure}

\begin{figure}
\includegraphics[width=0.9\textwidth{}]{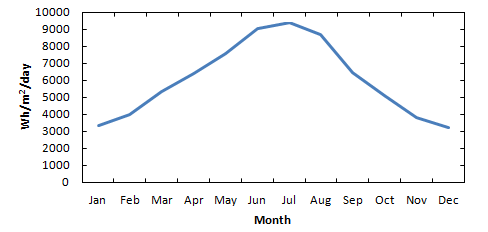}
\caption{DNI Direct Normal Irradiance (Wh/m$^2$/day)} 
\end{figure}
\clearpage 
\centering
\phantomsection


\addcontentsline{toc}{section}{\refname}
\begin{thebibliography}{99}
\bibitem{stazionemeteo:2011ug} 
\textbf {Tasselli, D. Ricci, S.}, ``Progetto Stazioni Meteoclimatiche e Sismologiche'', TS Corporation Srl - Dipartimento Meteorologia e Climatologia, 2014. 
\bibitem{Tasselli:2011ug} 
\textbf {Ente per le Nuove Tecnologie, l'Energia e l'Ambiente}, ``Tabella dei gradi/giorno dei Comuni italiani raggruppati per Regione e Provincia, Legge 26 agosto 1993, n. 412, allegato A'', 1 marzo 2011, p. 151.  
\bibitem{datimeteo:2013} 
\textbf {IlMeteo.it} ``Archivio Dati Meteo Taviano (LE)'', IlMeteo.it.
\bibitem{eumetsat:2013} 
\textbf {Eumetsat} ``Archivio Dati Meteo Taviano (LE)'', Eumetsat. 
\bibitem{atlantevento:2013} 
\textbf {Atlantedelvento} ``Archivio Dati Atlante Eolico'', http://www.atlanteeolico.it. 
\bibitem{ISPRA:2015} 
\textbf {ISPRA} ``Mappa Geologica di Taviano (LE) - '', ISPRA - Dipartimento Difesa del Suolo - Servizio Geologico d'Italia - Regione Puglia (2006), Foglio 223- http://www.isprambiente.it
\bibitem{RegionePuglia:2015} 
\textbf {WebGis} ``WebGIS  Carta Idrogeomorfologica della Puglia'', 2015
\bibitem{UE:2013}
\textbf{European Commission} ``Italy - Map of Solar Irradiation'', 2012
\bibitem{atlanteterra:1999}
\textbf{Garzanti} ``Atlante della Terra - Carta del Rischio Ambientale in Italia'', 1999
\bibitem{Zito G., Ruggiero L., Zuanni F.}
\textbf{Zito G., Ruggiero L., Zuanni F.} ``Aspetti meteoreologici e climatici della Puglia, atti 1°
workshop “clima, ambiente e territorio nel Mezzogiorno”, Taormina: 43-73', 1989
\end{thebibliography}
\end{document}